\documentclass[pra,twocolumn,aps,showpacs]{revtex4}
\usepackage[T1]{fontenc}
\usepackage[latin1]{inputenc}
\usepackage{bm}
\usepackage{amsmath}
\usepackage{amssymb}
\usepackage{mathrsfs}
\usepackage{latexsym}
\usepackage{amsfonts}
\usepackage{epsfig}
\usepackage{psfrag}
\newcommand{\ket}[1]{\left|#1\right>}
\newcommand{\bra}[1]{\left<#1\right|}

\newcommand{\bea}{\begin{eqnarray}}
\newcommand{\ea}{\end{eqnarray}}
\newcommand{\bmult}{\begin{multline}}
\newcommand{\emult}{\end{multline}}
\newcommand{\eea}{\end{eqnarray}}
\newcommand{\ord}{{\cal O}}
\newcommand{\sumint}[1]

\begin{document}

\title{Tunneling-induced damping of phase coherence revivals in deep 
optical lattices} 

\author{Uwe R. Fischer$^{1}$ and Ralf Sch\"utzhold$^{2}$}

\affiliation{$^{1}$Eberhard-Karls-Universit\"at T\"ubingen,
Institut f\"ur Theoretische Physik\\
Auf der Morgenstelle 14, D-72076 T\"ubingen, Germany
\\
$^{2}$Fachbereich Physik, Universit\"at Duisburg-Essen, 
D-47048 Duisburg, Germany}

\begin{abstract}
We consider phase coherence collapse and revival in deep optical lattices, 
and calculate within the Bose-Hubbard model the revival amplitude damping 
incurred by a finite tunneling coupling of the lattice wells 
(after sweeping from the superfluid to the Mott phase).  
Deriving scaling laws for the corresponding decay of first-order coherence 
revival in terms of filling factor, final lattice depth, and number of 
tunneling coupling partners, we estimate whether  revival-damping related 
to tunneling between sites can be or even has already been observed in 
experiment.  
\end{abstract} 


\maketitle 
{\em Introduction.}  
The idea of implementing the Bose-Hubbard model with cold atoms 
in optical lattices ultimately led to a clean experimental realization 
of the superfluid-Mott transition \cite{Jaksch,Greiner}. 
This in turn inspired 
many intriguing developments marrying condensed 
matter physics, quantum optics, and quantum information processing 
\cite{Bloch}. 
Concurrent with the superfluid-Mott transition, revivals of 
first-order phase coherence upon entering the Mott phase in a 3D 
optical lattice have been observed \cite{CollapseBEC}.
In the ideal case of a perfectly homogeneous lattice, 
the revivals 
would be exact, i.e., reach the same coherence amplitude as the initial 
superfluid state, provided the optical lattice wells were completely 
isolated from each other and from the environment (heat bath). 
The experiment \cite{CollapseBEC} however measures a significant damping
of the revival amplitude (so that approximately four to five revivals 
can be observed clearly).  
Here, the revival time $t_{\rm rev}$ is defined to be the time at which 
the first revival maximum of the long-range coherence 
$\langle\hat a^\dagger_\mu(t)\hat a_\sigma(t)\rangle$ 
occurs after the sweep to the Mott phase.  
%
There are basically four 
mechanisms which 
might induce the observed damping: 
$i)$ a finite tunneling coupling $J$ of the lattice wells,  
$ii)$ a finite coupling of the wells to the environment,  
and finally inhomogeneities in the lattice, such as
$iii)$ an external trap potential (inducing finite-size effects), and 
$iv)$ variations of the lattice depth and hence on-site repulsion 
$U$ from site to site (i.e., diagonal disorder). 
Here, we investigate the question of whether and how a finite 
tunneling coupling $J$ of lattice wells induces a damping of the 
revival signal, i.e., mechanism $i)$.
In contrast to the other effects, this mechanism is an intrinsic 
property of the system and survives in the homogeneous continuum limit.
It is also theoretically well under control and of fundamental interest, 
e.g., regarding the basic problem of equilibration, that is the question 
of whether and how fast the system approaches a local equilibrium state 
after the quench to the Mott phase. 
We point out, as 
these arguments already indicate, that the nature of  
revivals and their damping in an  optical lattice is fundamentally  
different from the single-well case where the particle-number 
dependence of the chemical potential plays the major role 
\cite{WrightWalls}.  
Our strategy 
is to calculate perturbatively the 
solution of the many-body problem on a general lattice, when the 
parameter $2\pi J/U$ is small but finite, starting from the  
exact number basis solution at $J=0$. 
Here $J$ gives a typical scale of the hopping amplitude, 
and $U$ is proportional to the contact interaction coupling. 
We note that within the presently employed pertubative 
approach, we cannot discuss the late-time behavior 
but can extract the decay of first-order coherence
from one given to the next revival cycle.  

{\em Tunneling-induced damping of revivals.} 
The Hamiltonian we employ is of the conventional single-band 
Bose-Hubbard type with on-site interactions 
\bea
\label{Hamiltonian}
\hat H
=
J 
M_{\bar\mu\bar\nu}\hat a_{\bar \mu}^\dagger\hat a_{\bar\nu}
+\frac{U}{2}
(\hat a_{\bar\mu}^\dagger)^2\hat a_{\bar\mu}^2
\,, \label{Hubbard}
\ea
where $\hat a_\mu^\dagger$ and $\hat a_\nu$ are bosonic 
creation and annihilation operators at lattice sites 
$\mu$ and $\nu$, respectively. 
We use the convention that summation 
is implicit over the equal Greek lattice site indices 
designated with an overbar. 
The matrix $M_{\mu\nu}$ describes the (possibly anisotropic) 
tunneling rates on an arbitrary lattice \cite{us}, and $U$ is the on-site 
repulsion. 
The full operator equation of motion in the final state with given 
$J$ then reads ($\hbar =1$)
\bea
i \frac{d\hat a_\mu}{dt} = J 
M_{\mu\bar\nu}\hat a_{\bar \nu}
+U\hat n_\mu\hat a_\mu \label{fullEOM}
\,,
\ea
where $\hat n_\mu=\hat a_\mu^\dagger\hat a_\mu$ counts the number of 
particles per site. 
The exact solution of the above operator equation, for decoupled wells, 
$J=0$, is given by 
\bea
\hat a^0_\mu(t)=\exp\{-iU\hat n_\mu t\}\hat a^0_\mu (0)
\,.
\ea
The revival time, where the argument of the exponential 
operator assumes integer multiples of $2\pi$, is  
$t_{\rm rev} = 2\pi/ U$ (for $J=0$). 
These coherence revivals exist because of the discrete spectrum 
of the filling operator $\hat n_\mu$, i.e., because of
the existence of particles, the ``granularity'' of matter.
Calculations for collapse and revival 
of first- and higher-order correlation functions
for the decoupled-well case $J=0$  
were performed in \cite{Bach}. 
We mention here that 
in the opposite limit of hard-core bosons ($U\rightarrow\infty$), 
hopping in an optical lattice from site to site, phase coherence 
collapse and revival may occur when an optical super-lattice is 
switched on \cite{Rigol}.  

We now proceed to calculate the perturbative solution of \eqref{fullEOM}
for $J/U$ small but finite.
Defining 
\bea
\hat a_\mu (t) = \exp\{-i U \hat n_\mu t\}\hat A_\mu (t), 
\ea
we extract the dominant (for $J/U\ll 1$) time-dependence from 
the full field operator $\hat a_\mu (t)$, so that 
$\hat A_\mu (0) = \hat a_\mu (0)$ and 
$\hat A_\mu (t_{\rm rev}) = \hat a_\mu (t_{\rm rev})$.
The equation of motion 
\bea 
i\partial_t \hat A_\mu = J 
M_{\mu\bar\nu} \exp\{iU(\hat n_\mu - \hat n_{\bar\nu}) t\} 
\hat A_{\bar\nu}  
\ea 
can be solved via an expansion into powers of the small parameter 
$J$ (analogous to response theory with a perturbation $\hat H_{\rm int}=
J M_{\bar\mu\bar\nu}\hat a_{\bar \mu}^\dagger\hat a_{\bar\nu}$), 
where the second-order solution reads 
\begin{multline}
\label{solutionJ^2}
\hat A_\mu (t) 
=  
\hat A_\mu (0) 
-i \int_0^t dt' J M_{\mu{\bar\nu}} e^{iU(\hat n_\mu - \hat n_{\bar\nu}) t'} 
\hat A_{\bar\nu} (0) 
\\
- \int_0^t dt'\int_0^{t'} dt'' J^2  M_{\mu{\bar\nu}}  M_{\bar\nu\bar\rho} 
e^{iU(\hat n_\mu - \hat n_{\bar\nu}) t'} 
e^{iU(\hat n_{\bar\nu} - \hat n_{\bar\rho}) 
t''} \hat A_{\bar\rho} (0) 
\\ 
+ \ord(J^3) 
\,.
\end{multline} 
Inserting the above second-order solution into the first-order 
correlation function evaluated at time $t_{\rm rev}$ yields 
\begin{multline}
\left\langle 
\hat A^\dagger_\mu (t_{\rm rev})  \hat A_\sigma (t_{\rm rev}) 
\right\rangle 
=
\left \langle 
\hat A^\dagger_\mu (0)  \hat A_\sigma (0) 
\right\rangle 
- \left(\frac{2\pi J} U\right)^2 \times 
\\ \times 
\left\{ 
\left\langle \hat A^\dagger_{\bar\nu} (0) 
 \delta(\hat n_\mu - \hat n_{\bar\nu}) 
\delta(\hat n_\sigma - \hat n_{\bar\rho}) 
\hat A_{\bar\rho} (0) 
\right\rangle M_{\mu\bar\nu} M_{{\bar\rho}\sigma} \right.
\\  
\left. + \frac12  \left\langle \hat A^\dagger_\mu (0) 
 \delta(\hat n_{\bar\nu} - \hat n_{\bar\rho}) 
\delta(\hat n_\sigma - \hat n_{\bar\rho}) 
\hat A_{\bar\nu} (0) 
\right\rangle   M_{\bar\nu\bar\rho} M_{\bar\rho\sigma} \right.
\\ 
\left. +\frac12 \left\langle \hat A^\dagger_{\bar\rho} (0) 
 \delta(\hat n_\mu - \hat n_{\bar\nu}) 
\delta(\hat n_{\bar\nu} - \hat n_{\bar\rho}) 
\hat A_\sigma (0) \right\rangle   
M_{\mu\bar\nu} M_{\bar\nu\bar\rho}  
\right\}
\\
+ \ord(J^3) 
\,,
\label{correlation} 
\end{multline} 
where $\delta (\hat n_\nu - \hat n_\rho) $ is to be understood
as an operator Kronecker delta, and the average is taken with
respect to the initial many-body quantum state. 
The above expression represents the full quantum result to second 
order in $J$, which is general insofar, within the single-band 
Bose-Hubbard model.  
This should be contrasted with the approach of \cite{Konotop}, 
which considers quantum collapse and revival in a two-dimensional (2D) 
respectively three-dimensional (3D) optical lattice, using 
resonant tunneling in the reduced dynamics of a two-mode (2D) 
or three-mode (3D) model.

The terms linear in $J$ generally vanish in the final result for the 
correlator \eqref{correlation}, which is to be expected because 
$\langle\hat A^\dagger_\mu(0)\hat A_\sigma(0)\rangle$  
assumes its maximum value ($\equiv n$ in the homogeneous case) 
in a coherent state.
Within linear response, the tunneling-induced 
damping therefore generally vanishes. 
%
We now use such a coherent state, more precisely a product of coherent 
states at each site, which has Poissonian
number statistics, to evaluate \eqref{correlation},  
\bea
|{\rm coh}\rangle 
= 
\prod_\mu |\alpha\rangle_\mu
= 
\prod_\mu e^{-|\alpha_\mu^2|/2} 
\sum_{n_\mu=0}^\infty 
\frac{\alpha_\mu^{n_\mu}}{\sqrt{n_\mu!}}   
|n_\mu\rangle \label{coh} 
\ea
where $|n_\mu\rangle$ are local number eigenstates,
$\hat n_\mu|n_\mu\rangle=|n_\mu\rangle n_\mu$
and $\hat a_\mu|\alpha\rangle_\mu=|\alpha\rangle_\mu\alpha_\mu$
with Poissonian distribution and average 
$\langle\hat n_\mu\rangle=|\alpha_\mu|^2$. 
The state \eqref{coh} is the factorized many-body state after a 
quasi-instantaneous sweep starting deep in the superfluid state 
\cite{us}. 
We assume homogeneity (no external trapping), $\alpha_\mu =\alpha$, 
$\langle\hat n_\mu\rangle = n$, and calculate the two different terms 
occurring in the correlator \eqref{correlation}. 
For the relevant case of long-range coherence (i.e., the two lattice sites 
$\mu$ and $\sigma$ are far apart and do not share neighbors, the result 
then being independent of the distance of the two sites), 
the first term gives  
\begin{multline}
\left\langle \hat A^\dagger_{\bar\nu} (0) 
 \delta(\hat n_\mu - \hat n_{\bar\nu}) 
\delta(\hat n_\sigma - \hat n_{\bar\rho}) 
\hat A_{\bar\rho} (0) \right\rangle M_{\mu\bar\nu} M_{\bar\rho\sigma}=
\\
= n \left( \sum_{k=0}^\infty p_{k,n}^2 \right)^2 D^2 
= n  e^{-4n} I_0^2 (2n) D^2
\,, 
\label{first} 
\end{multline}
where $D$ is the number of neighboring sites defined by 
$D=\sum_\nu M_{\mu\nu}$ for a fixed site $\mu$, 
with the $\nu$-sum running in the simplest case 
over only nearest neighbors to site $\mu$ 
(see however, discussion below of 
including next-nearest neighbors as well), 
and $I_0$ is modified Bessel function; the 
Poisson probabilities to find $k$ particles in a coherent 
state with the number average $n$ are denoted $p_{k,n} = e^{-n} n^k/k!$.
Similarly (for long-range coherence), the last two terms in 
\eqref{correlation} each give a contribution 
\begin{multline}
\frac12 \left\langle \hat A^\dagger_\mu (0) 
 \delta(\hat n_{\bar\nu} - \hat n_{\bar\rho}) 
\delta(\hat n_\sigma - \hat n_{\bar\rho}) 
\hat A_{\bar\nu} (0) 
\right\rangle   M_{\bar\nu\bar\rho} M_{\bar\rho\sigma} 
\\
= \frac12 n \left(\sum_{k=0}^\infty p_{k,n}^2\right)^2  D(D-1) 
+ \frac12 n  \sum_{k=0}^\infty p_{k,n}^3  D 
\\
= \frac{n}2 e^{-4n} I_0^2 (2n) D(D-1) \label{second} 
 + \frac{n}2 e^{-3n} F_{\{1,1\}}(n^3) D
\,,
\end{multline}
where $F_{\{1,1\}}(n^3)$ is a hypergeometric function.
The second term is generally negligible for $D\gg 1$, cf.
Fig.\,\ref{Contributions},   
because $\sum_{k=0}^\infty p_{k,n}^3 \sim 
(\sum_{k=0}^\infty p_{k,n}^2)^2 $ for the experimentally 
relevant range of $n=1\ldots 10$, with the  ratio of 
$\sum_{k=0}^\infty p_{k,n}^3$ to 
$(\sum_{k=0}^\infty p_{k,n}^2)^2 $
slowly decreasing with increasing $n$. 

Collecting the correlation functions from \eqref{first} and \eqref{second}, 
the total result for the decay of first order coherence 
at the instant of the first revival reads (see also Fig.\,\ref{Contributions})
\begin{multline}
\Delta (t_{\rm rev}) 
\equiv  
\frac{
\left \langle 
\hat A^\dagger_\mu (0) \hat A_\sigma (0) 
\right\rangle
-
\left\langle 
\hat A^\dagger_\mu (t_{\rm rev}) \hat A_\sigma (t_{\rm rev}) 
\right\rangle}n
=
\\
= 
\left(\frac{2\pi J} U\right)^2 
\left[ 
 D(2D-1) e^{-4n} I_0^2 (2n) \right.  
\left. + D e^{-3n} F_{\{1,1\}}(n^3) 
\right] 
.\\
\label{Delta}
\end{multline} 
This represents our major result for a coherent state with Poissonian 
statistics, i.e., after a sudden sweep from deep within the superfluid
phase, for which the original ground state \eqref{coh}
has no time to adjust.
\begin{center}
\begin{figure}[hbt]
\centerline{\epsfig{file=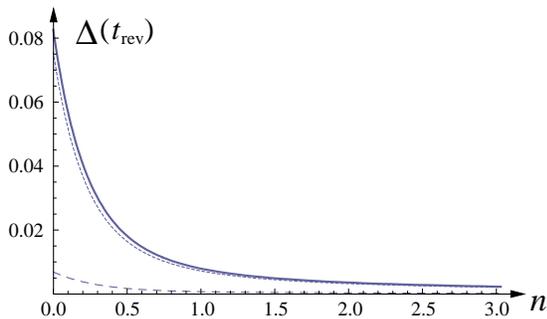,width=0.4\textwidth}}
\caption{\label{Contributions} 
The result \eqref{Delta} for the damping rate as a function of the filling $n$
(bold line), 
for $(2\pi J/U)^2 = 1.15\times 10^{-3}$ and $D=6$ \cite{CollapseBEC}. 
Dotted and dashed
lines represent the contributions of the $I^2_0$  and $F_{\{1,1\}}$
 terms in \eqref{Delta}, respectively.}
\end{figure}
\vspace*{-2.7em}
\end{center} 

{\em Numerical estimates and discussion.} 
We now provide a numerical estimate of \eqref{Delta} for the 
experiment \cite{CollapseBEC}, which was performed in a cubic 
3D optical lattice, with an initial depth $V_A = 8 E_r$ 
(corresponding to the superfluid phase), and a subsequent ``jump'' 
to a final depth $V_B = 22 E_r$ (deep into Mott phase).
In terms of the recoil energy $E_r = 2\pi^2/(m\lambda^2)$ determined by 
the laser wavelength $\lambda$, the tunneling coupling $J$ in 
sufficiently deep optical lattices can be estimated by 
$J/E_r = 8 (V/E_r)^{3/4} \exp[-2 \sqrt{V/E_r}]/\sqrt\pi$. 
Similarly, the on-site repulsion scales as 
$U/E_r = 4\sqrt{2\pi} (V/E_r)^{3/4} a_s/\lambda$,
where $a_s$ is the $s$-wave scattering length \cite{Boers}. 
One obtains for the perturbative parameter of our 
calculation 
\bea 
\left(\frac{2\pi J}{U}\right)^2
= 
\frac{8\lambda^2}{a_s^2} \exp 
\left[ - 4 \sqrt{\frac{V}{E_r}}
\right]
\,. 
\ea
Assuming, like for $^{87}\!$Rb, $a_s=$ 6\,nm, and a laser wavelength of 
$\lambda =$ 852\,nm, we have for $V=V_B= 22 E_r$ the value  
$(2\pi J/U)^2 = 1.15\times 10^{-3}$. 
With the number $D=6$ of nearest neighbors in a  3D cubic lattice 
and the experimental value \cite{CollapseBEC} of the average filling, 
$n=2.5$, we obtain $\Delta (t_{\rm rev}) = 2.8\times 10^{-3}$.  
Thus damping induced by tunneling to nearest neighbors, evaluated on the 
basis of a superfluid state with Poissonian statistics 
cannot directly explain the amount of damping observed in experiment 
\cite{CollapseBEC}, which is one to two orders of magnitude larger 
\cite{frequency}. 

The result \eqref{Delta} however represents a lower bound for 
tunneling-induced damping.
We now discuss various scenarios which would increase 
$\Delta(t_{\rm rev})$ above this lower bound. 
Firstly, \eqref{Delta} is the result 
obtained using the Poissonian superfluid state ($J\gg U$) 
as the initial unperturbed (zeroth-order) state. 
%
%
Assuming sub-Poissonian number statistics with a reduced number 
variance increases the expectation values of Kronecker symbols 
$\delta(\hat n_\mu - \hat n_\nu)$. 
In view of $\bra{\psi}\delta(\hat n_\mu - \hat n_\nu)\ket{\psi}\leq1$ 
for all states $\ket{\psi}$, we may obtain a crude upper estimate by just 
omitting all Kronecker symbols $\delta(\hat n_\mu - \hat n_\nu)$
in the expectation values. 
Inserting this upper bound, reasonable agreement with experiment 
can be reached, noting that $\sum_{k=0}^\infty p_{k,2.5}^3 \simeq 0.034$ 
and $(\sum_{k=0}^\infty p_{k,2.5}^2)^2 \simeq 0.0385$, 
which would then essentially be replaced by unity. 
Note, however, that actually realizing this upper bound would imply 
that the many-particle state approaches a product of Fock states with 
exactly $n$ particles at each site -- for which the coherence signal 
$\langle\hat A^\dagger_\mu \hat A_\sigma \rangle$ vanishes. 
Nevertheless, a state ``in-between'' the coherent state and 
the Mott state could still display non-vanishing long-range coherence 
(though below its maximum),  
$n>\langle\hat A^\dagger_\mu \hat A_\sigma \rangle>0$, 
whose revivals are more strongly damped 
than in the fully coherent case \eqref{Delta}. 
Actually, such an intermediate state with sub-Poissonian number statistics
is automatically created by a superfluid-Mott quench with a {\em finite}  
sweep rate \cite{us}.
The character of the final state (i.e., whether it is closer to the 
initial superfluid phase or the Mott state) depends on the time scale
of the quench in comparison with the (inverse) chemical potential, i.e., 
the relevant internal energy scale \cite{us}.  
From the revival time of $550\, \mu$sec \cite{CollapseBEC}, 
one obtains for $n=2.5$ an inverse chemical potential of 
$\mu^{-1} = (Un)^{-1} = 40\, \mu$sec.  
According to \cite{CollapseBEC}, the system is quenched in $50\, \mu$sec 
from superfluid to Mott.
The initial $V_0/E_r=8$ and final values $V_0/E_r=22$ imply that 
$J$ is quenched across approximately 3.7 $e$-foldings.
In order to compare the results with the calculation in \cite{us},
we assume an exponential decay $J(t)\propto \exp\{-\gamma t\}$
for simplicity.  
This implies that $1/\gamma\simeq 13.5\,\mu $sec, which is 
about a factor of three smaller than the inverse chemical potential of 
$\mu^{-1} = (Un)^{-1} = 40\, \mu$sec.  
Therefore, the sweep is rather non-adiabatic, which corresponds 
to a small adiabaticity parameter $\nu=Un/\gamma\approx1/3$
introduced in \cite{us}.
However, the sweep is not sudden -- for a sudden quench, 
the number fluctuations would essentially retain their 
initial value, $\langle\hat n^2\rangle-
\langle\hat n \rangle^2 = 
\langle\delta\hat n^2\rangle=n$ for a coherent state.
Due to the finite sweep rate $J(t)\propto \exp\{-\gamma t\}$
with $\nu=Un/\gamma\approx1/3$, though, the number variance 
is reduced by 60\%, i.e., $\langle\delta\hat n^2\rangle\approx0.4n$. 
Even though this result, derived in \cite{us} applies, strictly speaking, 
in the limit of large fillings $n\gg1$ only, one would expect a similar
sub-Poissonian statistics also for $n=2.5$, which would then increase 
the decay $\Delta(t_{\rm rev})$ in \eqref{Delta} significantly. 
An additional source for sub-Poissonian statistics could be the initial
state itself, if we do not start deep in the superfluid phase. 
Closer to the transition line (but still on the superfluid side), 
the number variance decreases, $\langle\delta\hat n^2\rangle<n$ 
already in the initial ground state. 
(In the experiment \cite{CollapseBEC}, merely 60\% of the atoms 
occupied the coherent condensate state initially.)

Another effect which potentially increases $\Delta(t_{\rm rev})$ is
to include couplings to next nearest neighbors, 
i.e., to effectively increase $D$. 
For the parameters of the experiment \cite{CollapseBEC}, 
the final next-nearest-neighbor couplings may be estimated as follows:
Within a harmonic approximation, the tunneling matrix elements 
are given by 
$JM_{\alpha\beta}\simeq J_{|\alpha-\beta|}\exp
\{-\pi^2\sqrt{V_0/E_r}(\alpha-\beta)^2/4\}$, 
where $|\alpha-\beta|$ 
is the distance of lattice points $\alpha$ and $\beta$ in units of 
the lattice spacing and $J_{|\alpha-\beta|}$ depends polynomially on 
$|\alpha-\beta|$ \cite{Illuminati,note}.
There are 20 next-nearest (diagonal) neighbors in a 3D cubic lattice 
in addition to the 6 nearest (straight) neighbors. 
Due to the exponential reduction of their individual contributions 
when quenching deep into Mott phase ($V_0/E_r =22$), 
their contribution gives about a factor of two for the damping 
$\Delta(t_{\rm rev})$. 
On the other hand, using only a moderately smaller final $V_0/E_r$ 
may make their total contribution to revival damping larger than 
that of the nearest neighbors.

{\em Concluding remarks.} 
In conclusion, we have derived a rigorous second-order result 
\eqref{correlation} for tunneling-induced damping of phase coherence 
revivals in optical lattices sufficiently deep inside 
the localized Mott phase $2\pi J\ll U$. 
Evaluating expression \eqref{correlation} for a coherent state,  
we obtained a lower bound for tunneling-induced damping \eqref{Delta},
which is too small to explain the experiment \cite{CollapseBEC}.
However, the incorporation of next-nearest (diagonal) neighbors 
as well as sub-Poissonian number statistics induced by the initial 
state and the finite sweep rate significantly enhance the tunneling-induced
damping $\Delta(t_{\rm rev})$ such that -- even though it does perhaps not 
fully reproduce the experiment \cite{CollapseBEC} -- it should 
constitute an observable fraction of the measured decay. 
Unfortunately, the precise value of the tunneling-induced damping 
of phase coherence cannot be derived from the information given in 
\cite{CollapseBEC} since part of the relevant input 
-- such as the exact dynamics $J(t)$ -- is missing. 
Thus, while the dephasing mechanisms $i)$ to $iv)$ discussed in the 
Introduction are qualitatively well understood, it is not possible to 
precisely disentangle their quantitative contributions from the data 
at hand.

While our prediction cannot be compared accurately to existing 
experimental results, a relatively modest modification of the 
parameters in an experiment like that of \cite{CollapseBEC}
will allow for its test. 
In particular, the exponential dependence of the tunneling-induced 
damping $\Delta(t_{\rm rev})\propto J^2$ on the laser
amplitude $\sqrt{V_0/E_r}$, implying 7.4 $e$-foldings for 
$(2\pi J/U)^2$ in the experiment \cite{CollapseBEC}, distinguishes 
(at low temperatures) this mechanism $i)$ from other sources 
like inhomogeneities due to external trapping $iii)$. 

We finally point out the importance of tunneling coupling 
for the equilibration of many-body states on the lattice. 
Nonequilibrium states of the Bose-Hubbard model were studied 
for various cases, e.g., quenching from the superfluid to the Mott side
\cite{us,Kollath}; from Mott to superfluid \cite{Cramer}; or for 
hard-core bosons in superlattices \cite{Rigol}.
While our approach is perturbative, yielding first-order coherence 
up to second order in $2\pi J/U$, it provides a first estimate on how 
important coupling of the wells can turn out to be for an equilibrium 
or non-equilibrium state to be established after a quench.  
Taking the decay of the first-order coherence as an indicator for 
locally approaching the equilibrium state, we conclude that 
equilibration occurs faster for many neighbors and for a larger 
tunneling rate (as one would expect) but also for sub-Poissonian 
number variance -- this is somewhat surprising, as 
one would generally expect that a state which is already 
closer to the Mott state decays slower. 

U.\,R.\,F. acknowledges support by the DFG under grant FI 690/3-1;  
R.\,S.~thanks A.~Pelster for discussions and 
acknowledges DFG support under grant SCHU~1557/1-3 
(Emmy-Noether Programme) and SCHU~1557/2-1 as well as SFB/TR12.

\end{document}